\documentclass[aps,preprint,manuscript,psfig,epsfig]{revtex4}
\usepackage{graphicx}

\begin{document}
\draft

\title{An Investigation of Hadronization Mechanism at $Z^{0}$ Factory}

\author{\sc Yi Jin\,$^{a}$, Zongguo Si\,$^{b}$, Qubing Xie$^b$ and Tao Yao\,$^b$}

\address{$^a$ Department of Physics, University of Jinan, Jinan Shandong, 250022, P. R. China
\\
$^b$ School of Physics, Shandong University, Jinan Shandong 250100, P. R. China
}

\begin{abstract}
We briefly review the hadronization pictures adopted in the LUND String Fragmentation Model(LSFM), Webber Cluster Fragmentation Model(WCFM) and Quark Combination Model(QCM), respectively. 
Predictions of hadron multiplicity, baryon to meson ratios and baryon-antibaryon flavor 
correlations, especially related to heavy hadrons at $Z^0$ factory obtained 
by LSFM and QCM are reported. \\
{\bf Keywords:}{~~$Z^0$ factory; Hadronization Model; 
baryon to meson ratio; $B\bar{B}$ 
correlation
}\\
{\bf PACS:}{~~12.15.Ji, 13.38.Dg, 24.10.Lx}
\end{abstract}

\maketitle

\renewcommand{\thefootnote}{\arabic{footnote}}

\section{Introduction}

\begin{figure}[h]
\begin{center}
\scalebox{0.9}{\includegraphics{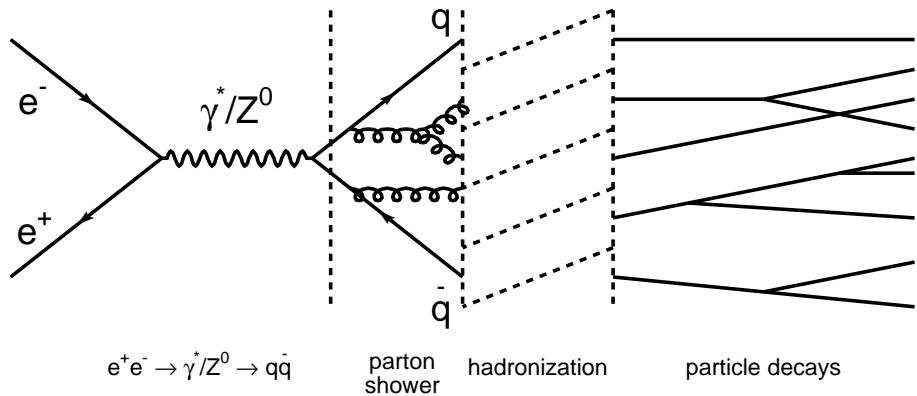}}
\caption{$e^{+}e^{-}$ annihilation to hadrons.} 
\label{hdpro}
\end{center}
\end{figure}

The hadronization mechanism is an important
but still unsolved problem up to now due to its nonperturbative nature.
It is recognized that the hadronization mechanism is universal in all kinds of
high energy reactions, {\it e.g}., $e^+e^-$ annihilation, and 
hadron(nuclear)-hadron(nuclear) collisions. Among these reactions,
$e^+e^-$ annihilation at high energies, especially at the 
$Z^0$ factory in the future, is best for studying the
hadronization mechanism, since all the final hadrons
come from primary ones, all of which are hadronization results.
The $e^+e^-\to \gamma^*/Z^0\to h's$ process is generally divided into
four phases (see Fig. \ref{hdpro}).
\begin{enumerate}
\item In the electro-weak phase, $e^{+}e^{-}$ pair converts into a primary
quark pair $q\bar{q}$ via virtual photon or $Z^{0}$. This phase is described by the electro-weak theory.

\item The perturbative phase describes the radiation of gluons
off the primary quarks, and the subsequent parton cascade due
to gluon splitting into quarks and gluons, and the gluon radiation
of secondary quarks. It is believed that perturbative QCD can describe this phase quantitatively.

\item In the hadronization phase, the quarks and gluons interact among themselves
and excite the vacuum in order to dress themselves into hadrons, that is, the confinement is `realized'.  Since this process belongs to the unsolved nonperturbative QCD, investigations employing various models will shed light on understanding this process.

\item In the fourth phase, unstable hadrons decay. This phase is usually
described by using experimental data.

\end{enumerate}
This paper focuses on step 3. The main method is to compare results of various hadronization models with the data, and the future $Z^0$ factory is very suitable for this purpose. The popular hadronization models at the market, Lund String Fragmentation Model(LSFM)\cite{sjostrand}, Webber Cluster Fragmentation Model(WCFM)\cite{Wolfram1980,Webber-cluster}, and
Quark Combination Model(QCM), succeed in explaining a lot of experimental data in $e^+e^-\to h's$ and $pp(\bar{p}) \to h's$ processes by adjusting corresponding parameters. Recently, the baryon
to meson ratio\cite{Adler:2003kg,Adams:2006wk} and constituent quark number scaling of elliptic flow $v_2$\cite{v2data} are measured at RHIC experiments, which do not favor the fragmentation model, while the QCM can explain these phenomena naturally\cite{Hwa:2002tu,Fries:2003vb,Greco:2003xt,AASDQCM}.QCM was first proposed by Annisovich and Bjorken {\it et al}. \cite{Anni}. 
It was famous for its simple picture and its successful prediction of the
percentage of vector mesons. One of its great merits is that it
treats the baryon and meson production in an uniform scheme, so it
describes the baryon production naturally. However, for a long period QCM is regarded as being ruled out, because the prediction for baryon-antibaryon($B\bar{B}$) rapidity correlation in $e^+e^-$  annihilation by Cerny's Monte Carlo Program, which was alleged to be based on QCM, has great discrepancies with the experimental results of TASSO collaboration\cite{TASSO}. But early in 1987, the $B\bar{B}$ phase space correlation from the naive QCM scheme was analyzed, which showed that there should not be such an inconsistency qualitatively. In the meantime, Quark Production Rule and Quark Combination Rule(the so-called `Shandong Quark Combination Model(SDQCM)') were developed. Then a serials of quantitative results obtained by SDQCM\cite{Xie1,Fang:1989hm,Si:1997zs} confirmed that SDQCM can naturally explain the $B\bar{B}$ short range rapidity correlation together with Baryon to Meson ratio when the multi-parton fragmentation is included. 
In order to understand the hadronization phenomena especially in heavy ion
collisions, it is necessary to study the hadronization mechanism in
detail in $e^+e^-$ annihilation once $Z^0$ factory is available. On the other hand,  the hadronization model serves as a bridge between the perturbative QCD and
experiments, so it is a 
very important tool for studying, {\it e.g.}, 
LHC physics. The properties of the light hadrons have been studied in our
previous works. Here we focus on investigating the production of
heavy hadrons({\it e.g}., $\Lambda_c, \ \Lambda_b,\ B_u$, {\it etc}.)  by LSFM
and SDQCM.


This paper is organized as follows: In section~\ref{ii}, we give a
brief introduction to the popular hadronization models, {\it i.e.}, LUND
String Fragmentation Model, Webber Cluster Fragmentation Model and
Quark Combination Model. Some numerical results are given in section
\ref{iii}. Finally, a short summary and outlook close the study.

\section{Hadronization Model}
\label{ii}
\subsection{LUND String Fragmentation Model}

String Fragmentation Model, first proposed by Artru and
Mennesser in 1974\cite{Artru:1974hr}, has been developed by the theory
physics group of Lund University since 1978, and corresponding
Monte-Carlo programs({\it e.g}., JETSET, PYTHIA, {\it etc}.)
are written. By now PYTHIA is one of the most
widely used generators describing the high-energy collisions.

The hadronization of $q\bar{q}$ color-singlet is the simplest
case for String Fragmentation. Lattice QCD supports a linear
confinement potential between color charges,{\it i.e.}, the energy stored
in the color dipole field increases linearly with the separation
between them. The assumption of linear confinement is the starting
point for the String Model. As the $q$ $\bar{q}$ move away in the 
opposite direction, the kinetic energy of the system changes into the
potential energy of the color string (or color flux tube). When the
potential increases to a certain extent the string will break by the
production of a new quark pair, and the production possibility is
given by the quantum mechanical tunnelling\cite{Glendenning:1983qq,Schenfeld:1990yu,Pavel:1990dz}
\begin{equation}\label{eq:tunneleffect}
    P_{q'(\bar{q}')}=e^{-\pi m_{T}^{2}/\kappa}=e^{-\pi m^{2}/\kappa}e^{-\pi p_{T}^{2}/\kappa}
\end{equation}
where $m_{T}$ is the transverse mass of $q'(\bar{q}')$ quark, and
$\kappa$, the string constant, denotes the potential energy per unit
length. Considering the assumption of no transverse excitation of
the string, the $p_{T}$ is locally compensated between $q'$ and
$\bar{q}'$. From Eq. (\ref{eq:tunneleffect}), one sees clearly that
the probability of the different flavors is about
$u:d:s:c\approx1:1:0.3:10^{-11}$. So $c,b$ quarks are not expected
to be produced through fragmentation, but only in the perturbative QCD
procedure.

When the new quark pair $q'\bar{q}'$ are excited out in vacuum, the
$q\bar{q}$ color string splits into $q\bar{q}'$ and $q'\bar{q}$ two
color-singlets. If the invariant mass of either of the singlet
system is large enough, further break will occur. The process ends
when all string pieces become exactly hadrons on their mass-shells.
This picture describes meson production naturally other than
the production of baryons. More complex diquark and popcorn mechanisms
have to be introduced to explain the baryon production.
One prominent feature of LSFM is that energy momentum and flavour
are conserved at each step of the fragmentation process.

\begin{figure}[htb]
\centering
\scalebox{0.3}{\includegraphics{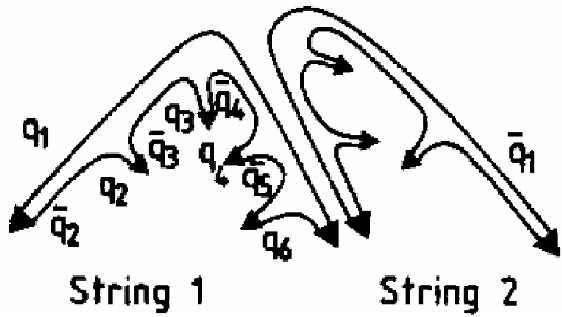}}
\caption{string fragmentation}
\label{figld}
\end{figure}

In $e^+e^-\to h's$ process at high energies, multi-parton states will
be produced at the end of perturbative phase, so that the multi-parton
fragmentation
has to be taken into account. The corresponding procedures have been
developed in ref. \cite{gust1982}. The  fragmentation picture of LUND model
is shown in Fig. \ref{figld}.

\subsection{Cluster Fragmentation Model}

\begin{figure}[htb]
\centering
\scalebox{0.3}{\includegraphics{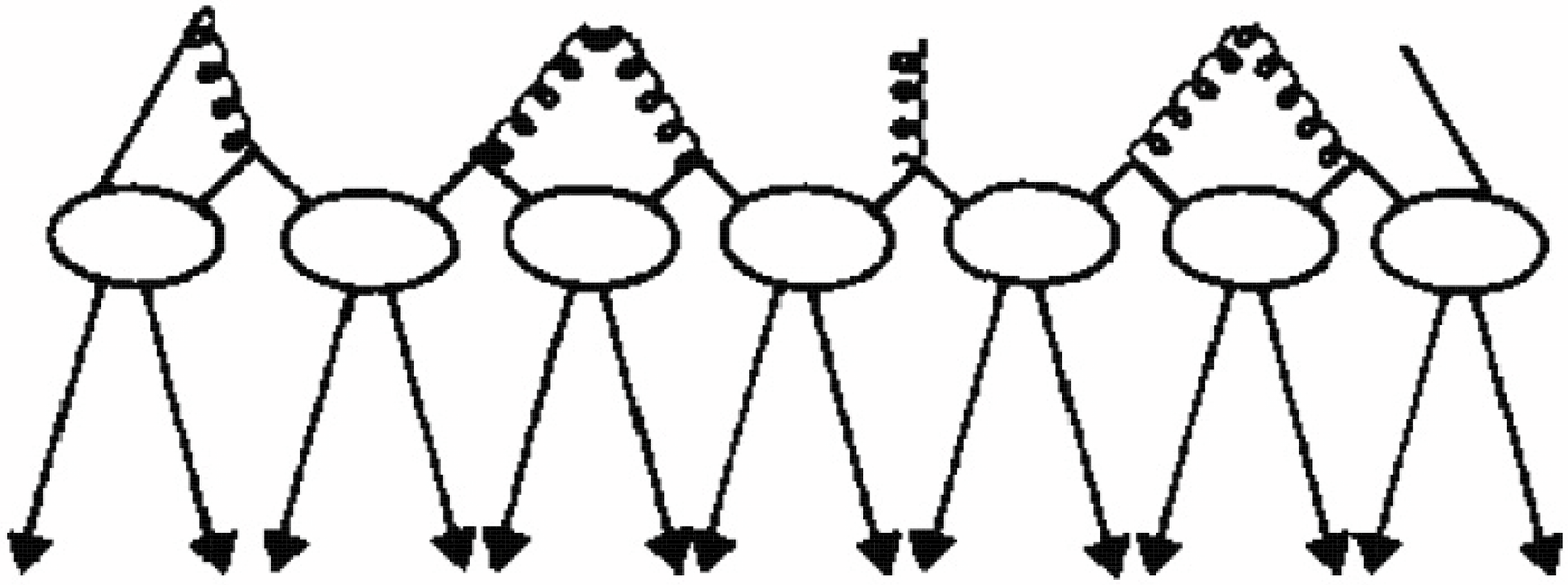}}
\caption{cluster fragmentation}
\label{figwebber}
\end{figure}

The Cluster Fragmentation Model has been proposed by Wolfram\cite{Wolfram1980} in 1980. Webber Cluster Fragmentation Model(WCFM)\cite{Webber-cluster} is the well-known example. It has three parts:
\begin{itemize}

\item The formation of color-singlet cluster. Gluons split into quark
pairs after parton shower: $g \rightarrow q\bar{q}$. Adjacent quark
and anti-quark from different gluons combine to form a color-singlet.

\item Cluster with large invariant mass fragments into smaller ones.

\item Little cluster decays into primary hadrons.{\it i.e.}, $Cluster
\rightarrow hadron1 + hadron2$.
\end{itemize}
Note that only two-body decay or fragmentation is adopted in the Webber cluster
fragmentation model. The corresponding
hadronization picture is local, universal and simple. 
Its fragmentation picture is shown in Fig. \ref{figwebber}.

\subsection{Quark Combination Model}

In the framework of SDQCM, Quark Production Rule(QPR),
Quark Combination Rule (QCR) are adopted to describe the
hadronization in a color singlet system, and then a simple Longitudinal Phase Space
Approximation~(LPSA) is used to obtain the momentum distribution for primary hadrons in its own system.
Finally this hadronization scheme is extended to the multi-parton states.
Here we briefly introduce the SDQCM and list the relevant equations(for
detail, see ref.~\cite{Xie2,Si:1997zs}).

\subsubsection  {QPR and QCR for $q\bar{q}$ system}
\label{m1}

In a color singlet system formed by $q\bar{q}$,
$N$ pairs of quarks can be produced by vacuum excitation via strong
interaction. We assume that $N$ satisfies Poisson Distribution:
\begin{equation}
\label{fir}
P(<N>,N-1) = {<N>^{N-1} \over (N-1)!} e^{-<N>}
\end{equation}
where $<N>$ is the average number of those quark pairs.
According to QPR, $<N>$ is given by
\begin{equation}
\label{fir1}
<N>=
\sqrt{\alpha^{2}+\beta(W-{{M}_{q}}-{{M}_{\bar{q}}}
+2\bar{m})} -\alpha-1, ~~~~~\\
\alpha=\beta\bar{m}-{{1} \over {4}}
\end{equation}
where $W$ is the invariant mass of the system, $\beta$ is a free
parameter, $\bar{m}$ is the average mass of newborn quarks,
and $M_{q}$ and $M_{\bar{q}}$ are the masses of endpoint quark
and anti-quark. Thus we have $N$ pairs of quarks according to
eqs.~(\ref{fir}),~(\ref{fir1}) (containing one primary quark pair).

When describing how quarks and antiquarks form hadrons, we find that
all kinds of hadronization models satisfy the near correlation in
rapidity more or less. Since there is no deep
understanding of the significance and the role of this, the near
rapidity correlation has not been
used sufficiently. In ref.~\cite{Xie1}, we have shown
that the nearest correlation in rapidity is in agreement with the
fundamental requirements of QCD, and determines QCR completely.
The rule guarantees that the combination of quarks across more
than two rapidity gaps never emerges and that $N$ quarks and $N$
antiquarks are exactly exhausted, which guarantees the unitarity (see ref.  \cite{Han:2009jw}). 
Given that the quarks and antiquarks are
stochastically arranged in rapidity space, each order can occur
with the same probability. Then the probability distribution for
$N$ quarks and $N$ antiquarks to combine into $M$ mesons, $B$
baryons and $B$ anti-baryons according to QCR is given by
\begin{equation}
\label{w1}
{X_{MB} =} {{2N(N!)^{2}(M+2B-1)!} \over {(2N)!M!(B!)^{2}}} 3^{M-1}
\delta_{N,M+3B}
\end{equation}
The average numbers of primary mesons $M(N)$ and baryons $B(N)$ are
\begin{equation}
\label{w2}
\left \{
\begin{array}{l}
M(N)=\sum\limits_{M,B}MX_{MB}(N) \\
B(N)=\sum\limits_{M,B}BX_{MB}(N)
\end{array}
\right.
\end{equation}
Approximately, in the combination for $N\geq 3$, $M(N)$ and
baryons $B(N)$ can be well parameterized as linear functions
of quark number $N$,
\begin{equation}
\label{w3}
\left\{
\begin{array}{l}
M(N)=aN+b\\
B(N)={{(1-a)} \over 3}N -{b \over 3}
\end{array}
\right.
\end{equation}
where $a=0.66$ and $b=0.56$. But for $N<3$, one has
\begin{equation}
\label{w4}
M(N)=N,~~~~B(N)=0~~~~for~~~ N<3
\end{equation}
So that, the production ratio of baryon to meson is obtained from
eqs.~(\ref{w2}) and (\ref{w3})
\begin{equation}
\label{w5}
{R_{B/M} =} {{(1-a)N -b} \over {3(aN+b)}}
\end{equation}
We see that SDQCM treats meson and
baryon formation uniformly, and there is no extra $ad~hoc$
mechanism and free
parameters for the baryon production. Here the $B/M$ ratio is
completely determined at a certain $N$, unlike in LSFM it
is completely uncertain and has to be adjusted by free
parameters, i.e, the ratio of diquark to quark ${qq}/q$.

\subsubsection {momentum distribution of primary
hadrons in the $q\bar{q}$ system}
\label{m2}
In order to give the momentum distribution of primary hadrons, each
phenomenological model must have some inputs. For example, in LSFM,
they use a symmetric longitudinal fragmentation function
\begin{equation}
\label{lu}
f(z) \propto {{(1-z)^a} \over {z}} exp(-b{m_{T}^{2} \over z})
\end{equation}
where $a$ and $b$ are two free parameters (and $a$ is
flavor dependent). In this paper, in order to give the momentum
distribution of primary hadrons produced according to QPR and QCR,
we simply adopt the widely used LPSA which is equivalent to the
constant distribution of rapidity. Hence a primary hadron $i$ is
uniformly distributed in the rapidity axis, and then its rapidity can be
written as
\begin{equation}
Y_{i}=Z+\xi_{i} Y, ~~~~
0 \leq \xi_{i} \leq 1,
\end{equation}
where $\xi_i$ is a random number;
$Z$ and $Y$ are two arguments, and can be determined by
energy-momentum conservation in such a color singlet system
\begin{equation}
{\sum \limits_{i=1}^{H}} E_{i} = W, ~~~~
{\sum \limits_{i=1}^{H}} P_{Li} = 0,
\end{equation}
where $E_{i}$ and $P_{Li}$ denote the energy and the longitudinal
momentum of the $i$th primary hadron respectively, obtained by
\begin {equation}
\left \{
\begin{array}{ll}
E_{i} = m _{Ti} {{exp(Y_{i}) + exp(-Y_{i})} \over {2}}\\
P_{Li}= m _{Ti} {{exp(Y_{i}) - exp(-Y_{i})} \over {2}}
\end{array}
\right.
\end{equation}
where $m_{Ti}$ is given by
\begin{equation}
m _{Ti} = \sqrt{m_{i}^2 + {\stackrel {\rightarrow} P_{Ti}}^{2}}
\end{equation}
where $m_{i}$ is the mass of the $i$th primary hadron,
and $\stackrel {\rightarrow} P_{Ti}$ obeys the distribution
\begin{equation}
\label{last}
f({\stackrel{\rightarrow} P_{T1}},\ldots,
{\stackrel{\rightarrow}P_{TH}}) \propto {\prod \limits_{i=1}^{H}}
exp(-{{{\stackrel {\rightarrow} P_{Ti}}^{2}} \over {\sigma^{2}}})
\delta ({\sum \limits_{i=1}^{H}} {\stackrel {\rightarrow} P_{Ti}})
\end{equation}
In this paper, we set $\sigma=0.2~GeV$. Eq.~(\ref{last}) is
just what LSFM uses.

Note that LPSA or the constant rapidity distribution is rather
naive, but it
is convenient for us to study the correlations without introducing
many complicating parameters.

\subsubsection {hadronization of a multi-parton state}
\label{m3}
\begin{figure}[htb]
\centering
\scalebox{0.6}{\includegraphics{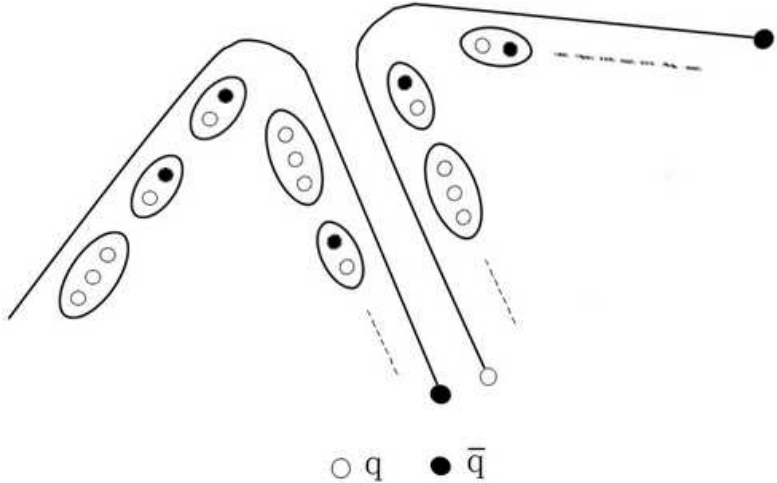}}
\caption{quark combination}
\label{qcm1a}
\end{figure}

At the end of parton showering, a final multi-parton state
will start to hadronize. To connect the final multi-parton state
with SDQCM, we adopt a simple treatment assumed in WCFM, {\it i.e.} before
hadronization, each gluon at last splits into a $q'\bar{q'}$
pair, the $q'$ and $\bar{q'}$ carry one half of the gluon
momentum, and each of them forms a color singlet  with their
counterpart antiquark and quark in their neighbourhood, respectively.
Now take the three parton state $q\bar{q}g$ as an example to
illustrate the hadronization of a multi-parton state. Denote the
4-momenta for $q$, $\bar{q}$, $g$ as
\begin{equation}
P_{1}  = ( E_{q} , {\stackrel {\rightarrow} P_{q}} ), ~~
P_{2}  = ( E_{g} , {\stackrel {\rightarrow} P_{g}} ), ~~
P_{3}  =(E_{\bar{q}},{\stackrel{\rightarrow}P_{\bar{q}}}).
\end{equation}
Before hadronization, the gluon splits into a $q'\bar{q'}$
pair and the  $q'$ and $\bar{q'}$ carry one half of the
gluon momentum, and the $q\bar{q}g$ system forms two
color singlet subsystems $q\bar{q'}$ and $q'\bar{q}$.
The invariant masses of the subsystems are
\begin{equation}
\left \{
\begin{array}{lll}
W_{q{\bar{q'}}} =& \sqrt {( P_{1} + {{P_{2}} \over {2}})^{2}}=
&\sqrt{(E_{q}+{{E_{g}}\over{2}})^{2}-({\stackrel{\rightarrow}P_{q}}+
{{\stackrel {\rightarrow} P_{g}} \over {2}})^{2}}\\
W_{{q'}{\bar{q}}} =&\sqrt {( P_{3} + {{P_{2}} \over {2}})^{2}} =
&\sqrt{(E_{\bar{q}}+{{E_{g}}\over{2}})^{2}-
({\stackrel{\rightarrow}P_{\bar{q}}}+
{{\stackrel {\rightarrow} P_{g}} \over {2}})^{2}}
\end{array}
\right.
\end{equation}
As was commonly argued by Sj\"{o}strand and Khoze
recently\cite{khoze}, the confinement effects should lead to a
subdivision of the full $q\bar{q}$ system into color singlet
subsystems with screened interactions between these subsystems
$q\bar{q'}$ and $q'\bar{q}$. Hence SDQCM can be applied
independently to each color singlet subsystem,{\it i.e.}, we can apply
the equations in the former two subsections to each subsystem, and
obtain the momentum distribution for the primary hadrons in their
own center-of-mass system. Then after Lorentz transformation, the
momentum distribution of the  primary hadrons in laboratory frame is 
given. This treatment can be extended to a general multi-parton
state, for example, see Fig. \ref{qcm1a}. 

Obviously, when the emitting gluon is soft or collinear with the
direction of $q$ or $\bar{q}$, $q\bar{q}g$ cannot be
distinguished from $q\bar{q}$ and $W_{q\bar{q}'}$ or
$W_{q'\bar{q}}$ is too small for hadronization. For the avoidance of these situations, a cut-off mass $M_{min}$ has to be introduced. Here
$M_{min}$ is a free parameter in the perturbative phase.
Its value and energy dependence is theoretically uncertain.
The physical assumption is that $M_{min}$ is
independent of energy\cite{Mattig,sjostrand}.

\section{Results and Discussion}\label{iii}

In this paper, we study the hadronization process at $Z^0$ 
factory by LSFM and SDQCM. We focus on investigating the properties of the
heavy hadrons, {\it e.g}., the ratio of  baryon to meson and the 
baryon-antibaryon 
correlations. For LSFM, 
we use the default parameter values
in PYTHIA, while for SDQCM, we adopt the 
parameters used in ref. \cite{Si:1997zs} which fit data quite well.

\begin{figure}[h]
\begin{center}
\scalebox{0.7}{\includegraphics{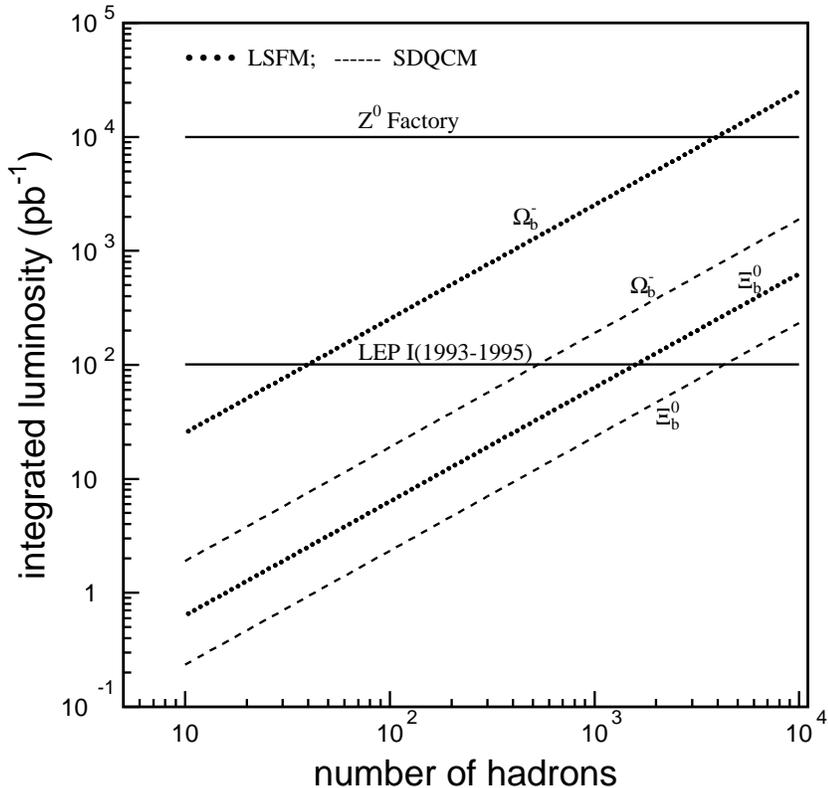}}
\caption{The relation between integrated luminosity and the production number of heavy bayons at $Z^0$ energy. The lower solid line stands for the total luminosity of 
LEP I from 1993 to 1995 at energies around the $Z^0$ \cite{Abbiendi:1999zx}.}
\label{lumz0}
\end{center}
\end{figure}

The final hadron multiplicity 
at high energy $e^+e^-$ reactions is always an important topic.  
We list the predictions of final hadron multiplicities at $Z^0$ 
factory($\sqrt{S}=91.2GeV$) by LSFM and SDQCM in table I. 
The results of LSFM are obtained by running PYTHIA6.4 with 
the default parameters. The experimental data are from ref. \cite{pdg}. 
It is found that the predictions of LSFM and those of SDQCM are consistent with most of the experimental data. 
However, the multiplicity of the heavy baryons, {\it  eg}., $\Xi_b$,
$\Sigma_b,~ \Omega_b$, are still not measured at $Z^0$ energy, and 
the corresponding theoretical predictions obtained by LSFM and SDQCM are quite different. Therefore it is important to measure their production rates at $Z^0$ pole in order to discriminate different hadronization 
mechanisms. At the $Z^0$ energy, we 
study the relation between the integrated luminosity and 
the  producing number of the heavy baryons. The corresponding results are 
displayed in Fig.\ref{lumz0}. 
It is obvious that for the LEP I, the production number for $\Xi_b$ predicted by
LSFM(SDQCM) is about $10^3$, and that for $\Omega_b$ is several tens(hundreds). 
In order to study the heavy baryon production mechanism with enough precision, 
the integrated luminosity should be increased as large as possible. For example,
if the integrated luminosity of $Z^0$ factory can reach about $10^4 pb^{-1}$, the produced number of $\Omega_b$ may reach several thousands(ten thousands) according to the prediction of LSFM(SDQCM), which makes it possible to study the properties of the heavy baryons and to test the hadronization models with higher precision.

\begin{table}[htb]
\label{multiplicities}
\caption{Results for average hadron multiplicities at $Z^0$ factory. 
The experimental data are from ref. \cite{pdg}.}
\begin{center}
\begin{tabular}{l c c c}
\hline
~Particle~~~~ & ~~~~EXP DATA~~~~ & ~~~~LSFM~~~~ & ~SDQCM~~\\
\hline
~~$\pi^+$  & 17.02  $\pm$0.19    & 17.125 & 17.766\\
~~$\pi^0$  &   9.42  $\pm$0.32    &   9.696 &  9.633\\
~~$K^+$    &  2.228 $\pm$0.059  &   2.312 &  2.145\\
~~$K^0$    &  2.049 $\pm$0.026  &   2.079 &  1.758\\
~~$\eta$    &  1.049 $\pm$0.080  &   1.013 &  0.787\\
~~$D^+$    &  0.175 $\pm$0.016  &   0.166 &  0.217\\
~~$D^0$    &  0.454 $\pm$0.030  &   0.495 &  0.476\\
~~$B^+$    &  0.178 $\pm$0.006  &   0.174 &  0.182\\
~~$B^0$    &  0.165 $\pm$0.026  &   0.173 &  0.182\\
~~$B_s^0$ &  0.057 $\pm$0.013  &   0.052 &  0.053\\
~~$\omega$    &  1.016 $\pm$0.026  &   1.369 &  1.809\\
~~$\rho^0$    &  1.231 $\pm$0.065  &   1.524 &  1.842\\
~~$K^{*+}$ &  0.715 $\pm$0.059  &   1.112 &  1.043\\
~~$K^{*0}$ &  0.738 $\pm$0.024  &   1.107 &  1.012\\
~~$D^{*+}$ &  0.1937 $\pm$0.0057  &   0.2399 &  0.2547\\
~~$D^{*0}$ &                          ---  &   0.2407 &  0.2603\\
~~$p$          & 1.050  $\pm$0.032    & 1.223 & 0.928\\
~~$\Lambda$ & 0.3915  $\pm$0.0065    & 0.3930 & 0.4368\\
~~$\Sigma^0$  & 0.076  $\pm$0.011    & 0.075 & 0.135\\
~~$\Sigma^-$  & 0.081  $\pm$0.010    & 0.069 & 0.114\\
~~$\Sigma^+$  & 0.107  $\pm$0.011    & 0.074 & 0.122\\
~~$\Lambda_c^+$ & 0.078  $\pm$0.017    & 0.060 & 0.076\\
~~$\Lambda_b^0$ & 0.031  $\pm$0.016    & 0.035 & 0.044\\
~~$\Sigma_c^0$  & ---                         & 0.0017 & 0.0073\\
~~$\Sigma_b^0$  & ---                         & 0.0019 & 0.0102\\
~~$\Xi_b^0$  & ---                         & 0.0024 & 0.0065\\
~~$\Omega_b^-$  & ---                         & 0.00006 & 0.0008\\
\hline
\end{tabular}
\end{center}
\end{table}

\begin{figure}[h]
\begin{center}
\scalebox{0.6}{\includegraphics{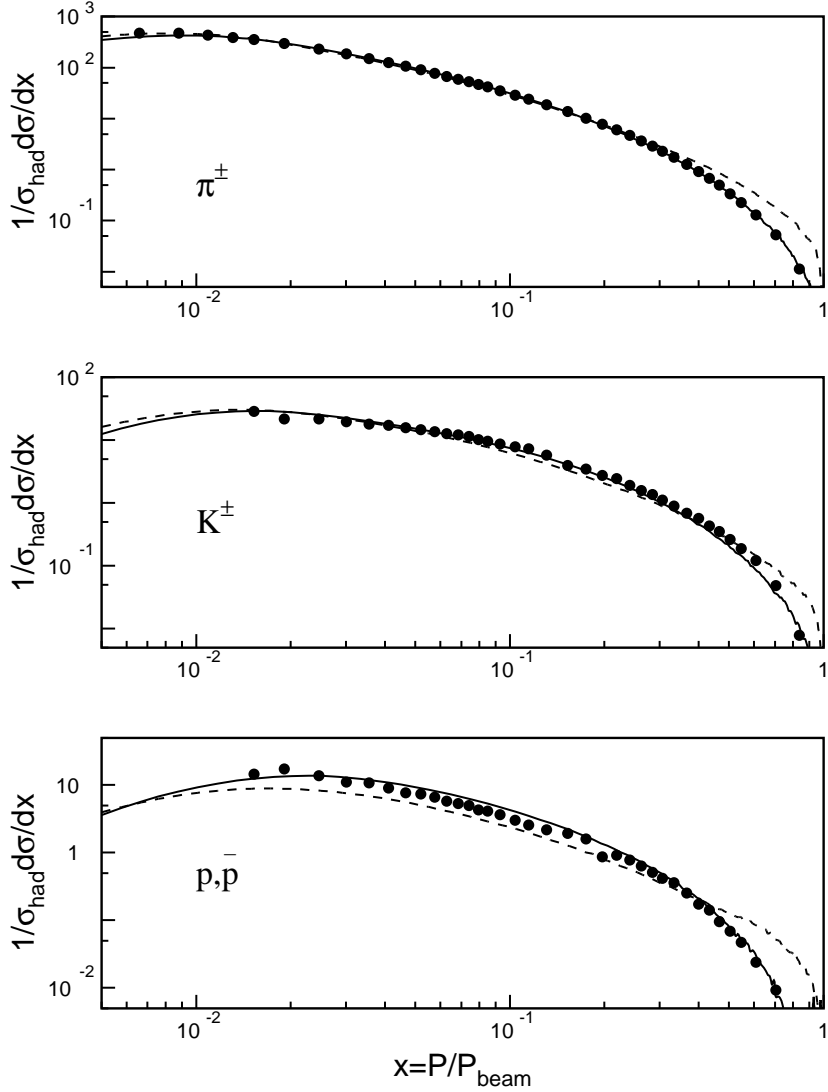}}
\caption{The momenta spectrum of $\pi^{\pm}$, $K^{\pm}$ 
and $p\bar{p}$ at $Z^0$ factory. Solid(dashed) line is for LSFM(SDQCM), and here $P_{beam}=\sqrt{S}/2$. The experimental data is from ref. \cite{OPAL}.}
\label{ppild}
\end{center}
\end{figure}

We also show the results for the momentum distribution 
of $\pi^{\pm}$, $K^{\pm}$ and $p\bar{p}$ in Fig.\ref{ppild}. 
The results of LSFM are more consistent than SDQCM. 
This is easy to understand, since SDQCM does not include any complicated inputs for the hadron momentum distributions to be improved in the future.

The following focuses on studying the baryon to meson ratio and 
$B\bar{B}$ correlations which reflect the hadronization mechanism more directly. In LSFM, the baryon to meson ratios can be tuned by the free parameters,
{\it e.g}., $qq/q$, $BM\bar{B}/(B\bar{B}+BM\bar{B})$. 
SDQCM describes the baryon and meson production in the uniform scheme, so that these ratios are obtained naturally. Some of these ratios are listed in table II at $Z^0$ factory. One can notice that both LSFM and SDQCM can explain the present data. In $e^+e^-$ annihilations, most baryons are produced directly, even if
they are the decay products. They carry the main properties of their mother particles. 

\begin{table}[htb]
\label{bmratio}
\caption{Results for baryon to meson ratios at $Z^0$ factory. The experimental data are from  ref. \cite{pdg}.}
\begin{center}
\begin{tabular}{l c c c}
\hline
~~~~~~~~~ & ~~~~EXP DATA~~~~ & ~~~~LSFM~~~~ & ~SDQCM~~\\
\hline
~~$\frac{\Lambda_b^0}{B^0}$  & 0.188  $\pm$0.101  & 0.201 & 0.239\\
~~$\frac{\Lambda_b^0}{B^+}$  & 0.174  $\pm$0.090  & 0.200 &  0.240\\
~~$\frac{\Lambda_c^+}{D^0}$  &  0.172 $\pm$0.039  & 0.121 &  0.160\\
~~$\frac{\Lambda_c^+}{D^+}$  &  0.446 $\pm$0.105  & 0.360 &  0.351\\
\hline
\end{tabular}
\end{center}
\end{table}

Studying the properties of baryons, especially the $B\bar{B}$ flavor correlations, 
is helpful to reveal the hadronic mechanism. 
Here the $B\bar{B}$ flavor correlation strength is defined as
$$R_{B\bar{B}}=N\frac{n_{pair}}{n_B+n_{\bar{B}}}$$
where $n_{pair}$ is the number of the $B\bar{B}$ pairs and 
$N_{B}(N_{\bar{B}})$ is the baryon(antibaryon) number. 
$N=2(1)$ in the case of $B=\bar{B}$($B\neq \bar{B}$)\cite{OPAL}.
The predictions of baryon antibaryon flavor correlation strength by SDQCM and LSFM are listed  in table III together with the corresponding OPAL data\cite{OPAL}. The understanding of hadronization mechanism related to the heavy hadron production requires more precise measurements at the future $Z^0$ factory. 

\begin{table}[htb]
\label{correlation}
\caption{Predictions of baryon antibaryon flavor correlations by SDQCM and 
LSFM, and the experimental results are from OPAL collaboration\cite{OPAL}.}
\begin{center}
\begin{tabular}{c c c c}
\hline
~~~~~~~~~ & ~~~~EXP DATA~~~~ & ~~~~LSFM~~~~ & ~SDQCM~~\\
\hline
~~$\Lambda\bar{\Lambda}$  & 0.49  $\pm$0.06    & 0.38 & 0.48\\
~~$\Xi^-\bar{\Xi}^+$  &   0.04  $\pm$0.06    &   0.14 &  0.15\\
~~$\Xi^-\bar{\Lambda}+\bar{\Xi}^+\Lambda$    &  0.463 $\pm$0.099  &  
0.510 &  0.538\\
~~$\Lambda_b^0\bar{\Lambda}_b^0$  & ---    & 0.08 & 0.12\\
\hline
\end{tabular}
\end{center}
\end{table}

\section{Summary and Outlook}
In the $e^+e^-\to \gamma ^\ast /Z^0\to h'$ process, the 
final particles have no relations with the structure of the initial particles, 
so it is convenient to study the hadronization mechanism. 
Above all, such studies help to investigate the  
multi-production processes in other high energy 
reactions({\it e.g.}, at RHIC energy).
In this paper, we report the predictions especially related to heavy hadrons,
for example, baryon to meson ratio and $B\bar{B}$ 
flavor correlations obtained by LSFM and SDQCM in the hadronic $Z^0$ decay. 
These results show that the future experiments at $Z^0$ factory to test hadronization mechanism, especially rare hadronized events, such as the doubly heavy baryon production, are significant.

Though many studies have been made for the $e^+e^-\to h's$ process, still a lot of unclear but important topics are worth studying, such as the color 
connection effects\cite{xsl}. The color connection among final partons plays a 
crucial role in the surface between PQCD phase and 
the hadronization one(see Fig. \ref{hdpro}). 
Different choices of the color connections of the final
multi-parton system leads to different hadronization results. 
Recently, the doubly heavy baryon({\it e.g.}, $\Xi_{cc}$) production has been
investigated in different reactions\cite{ms}. In $e^+e^-$ annihilation,
the doubly heavy baryon production
reveals a special kind of color connection of the final parton system 
$q_1\bar{q}_1 q_2\bar{q}_2+ng$($n\leq 0$). This kind of color connection 
has not been considered in the popular hadronization models. 
In ref. \cite{hanwei}, the hadronization effects
of this kind of color connection are investigated in an extremely limited case 
of $e^+e^-\to c\bar{c}q\bar{q} \to h's$ process  at $B$ 
factory energies. There {\it cq} and $\bar{c}\bar{q}$ form 
diquark-antidiquark pair, which is treated as a color singlet 
string and fragments into hadrons in the framework of LSFM.
However due to the phase space limitation, the hadronization effect induced by
diquark pair fragmentation is small. Fortunately, in the future 
$Z^0$ factory, a large number of events such as 
$e^+e^-\to Z^0\to \Xi_{cc} \bar{c} \bar{c} X$ will be 
produced\cite{zhangzx}, so that the correlated physical observables 
can be measured with higher statistics.
As a result, the hadronization mechanism and the related non-perturbative nature
can be well studied at $Z^0$ pole energy.


\section*{Acknowledgements}
This work is supported in part by NSFC, Natural Science 
Foundation of Shandong Province(ZR2009AM001), and Doctoral Science Foundation 
of University of Jinan(B0527). The authors would like to 
thank Prof. S. Y. Li for his kindly helpful discussions.

\end{document}